\documentclass[%
 reprint,
 amsmath,amssymb,
 aps,superscriptaddress,
]{revtex4-1}

\usepackage{color}
\usepackage{amsmath}
\usepackage{natbib}
\usepackage[export]{adjustbox}
\usepackage{graphicx}
\usepackage{dcolumn}
\usepackage{bm}
\usepackage{subfigure}
\usepackage{tikz}
\usetikzlibrary{arrows}
\usepackage{multirow}

\tikzset{font={\fontsize{5pt}{5}\selectfont}}

\begin{document}

\title{The role of dipole-forbidden autoionizing resonances in non-resonant one-color two-photon single ionization of N$_2$}

\author{Kirk A. Larsen}
\email{KLarsen@lbl.gov}
\affiliation{%
 Graduate Group in Applied Science and Technology, University of California, Berkeley, CA 94720, USA}
\affiliation{%
 Chemical Sciences Division, Lawrence Berkeley National Laboratory, Berkeley, CA 94720, USA}%

\author{Roger Y. Bello}
\email{RYBelloromero@lbl.gov}
\affiliation{%
 Chemical Sciences Division, Lawrence Berkeley National Laboratory, Berkeley, CA 94720, USA}%
 
\author{Robert R. Lucchese}
\affiliation{%
 Chemical Sciences Division, Lawrence Berkeley National Laboratory, Berkeley, CA 94720, USA}%
 
\author{Thomas N. Rescigno}
\affiliation{%
 Chemical Sciences Division, Lawrence Berkeley National Laboratory, Berkeley, CA 94720, USA}%
 
\author{C. William McCurdy}
\affiliation{%
 Chemical Sciences Division, Lawrence Berkeley National Laboratory, Berkeley, CA 94720, USA}%
\affiliation{%
 Department of Chemistry, University of California, Davis, CA 95616, USA}%
 
\author{Daniel S. Slaughter}
\affiliation{%
 Chemical Sciences Division, Lawrence Berkeley National Laboratory, Berkeley, CA 94720, USA}%
 
\author{Thorsten Weber}
\email{TWeber@lbl.gov}
\affiliation{%
 Chemical Sciences Division, Lawrence Berkeley National Laboratory, Berkeley, CA 94720, USA}%

\date{\today}

\begin{abstract}
We present an experimental and theoretical energy- and angle-resolved study on the photoionization dynamics of non-resonant one-color two-photon single valence ionization of neutral N$_2$ molecules. Using 9.3~eV photons produced via high harmonic generation and a 3-D momentum imaging spectrometer, we detect the photoelectrons and ions produced from one-color two-photon ionization in coincidence. Photoionization of N$_2$ populates the X $^2\Sigma^+_g$, A $^2\Pi_u$, and B $^2\Sigma^+_u$ ionic states of N$_2^+$, where the photoelectron angular distributions associated with the X $^2\Sigma^+_g$ and A $^2\Pi_u$ states both vary with changes in photoelectron kinetic energy of only a few hundred meV. We attribute the rapid evolution in the photoelectron angular distributions to the excitation and decay of dipole-forbidden autoionizing resonances that belong to series of different symmetries, all of which are members of the Hopfield series, and compete with the direct two-photon single ionization.
\end{abstract}

\pacs{Valid PACS appear here}

\maketitle

\section{\label{sec:level1}Introduction}

The photoionization continuum of molecular nitrogen posses both one- and two-electron resonances, i.e. shape resonances and autoionizing states, which can strongly influence the behavior of ejected photoelectrons. The observable that is most sensitive to these resonance effects is the energy-resolved photoelectron angular distribution (PAD). In the case of single photon ionization these resonance effects have been studied extensively both experimentally and theoretically \cite{Lucchese0,Lucchese3,Lucchese4,Lucchese5,Ilchen,Plummer,Dehmer,Dehmer1,Raoult,Ogawa,Huber,Klinker,Yencha,Plesiat,Plesiat1}. However, due to its complexity, non-resonant two-photon ionization has received far less attention, while it provides the opportunity to observe the effect of discrete continuum states on the angular distributions of the photoelectrons and ions that are symmetry- and/or parity-forbidden in single photon excitation schemes~\cite{Larsen1}. Moreover, two-photon absorption resulting in single ionization is particularly sensitive to resonance and electron-electron correlation effects \cite{Larsen2}.

The presence of continuum-embedded discrete states can lead to interference effects, a paradigmatic example being the Fano resonance, where the scattering amplitudes of a resonant (discrete) and background (continuum) process interfere to give rise to the classic asymmetric line shape \cite{Fano}. In photoionization, continuum states that decay via autoionization can provide an indirect channel to target states that can interfere with the direct ionization channel to those very same states \cite{Bello,Vrakking,Pfeifer,Sansone,Chang}. This effect can be observed in a PAD \cite{Vrakking,OKeeffe,Martin}, where the amplitudes and phases of the angular momentum components of the photoelectron scattering wave function from direct ionization interfere with those of the electron originating from autoionization.

The dipole oscillator strength of N$_2$ across most of the VUV photon energy range is quite weak, as the states that lie below $\sim$12~eV are dipole-forbidden transitions, because of either spin- or parity-conservation. Within the ionization continuum and below the B $^2\Sigma^+_u$ ionic state (situated at 18.74~eV in the Franck-Condon (FC) region) lies the Hopfield series of N$_2$ \cite{Hopfield}, a progression of narrow resonances emerging from continuum molecular Rydberg states converging to the B $^2\Sigma^+_u$ ionic state. Such quasi-bound states can decay through autoionization, where these two-electron resonances provide an indirect channel to the X $^2\Sigma^+_g$ and A $^2\Pi_u$ ionic states, along with the direct ionization channel to those very same target states. By choosing a suitable photon energy, these autoionizing states can thus be populated and studied via non-resonant one-color two-photon ionization.

In this work we present experimental and theoretical results on energy- and angle-resolved Non-resonant One-color Two-Photon Single Ionization (NOTPSI) of molecular nitrogen using 3-D momentum imaging, where the photoelectron and molecular cation are measured in coincidence. Using a 400~nm driving field, we produce and select VUV photons with an energy of 9.3~eV via High Harmonic Generation (HHG), which are then used to singly ionize the target via two-photon ionization. We observe that the energy-resolved PADs associated with the X $^2\Sigma^+_g$ and A $^2\Pi_u$ ionic states both vary appreciably with small changes in the photoelectron kinetic energy. We ascribe these effects to the excitation and decay of dipole-forbidden autoionizing resonances belonging to series possessing different total symmetries, which, in addition to the direct ionization channel, provide an indirect pathway to the target states.

\section{\label{sec:level2}Experiment}

The valence photoionization dynamics in neutral N$_2$ were investigated using the COLd Target Recoil Ion Momentum Spectroscopy (COLTRIMS) technique~\cite{dornerColdTargetRecoil2000,Ullrich,Jahnke,Sturm}, where the photoelectrons and cations produced by NOTPSI are collected with full 4$\pi$ solid angle, and their 3-D momenta are measured in coincidence on an event-by-event basis. The charged particles are guided by parallel DC electric and magnetic fields (11.34~V/cm, 4.0~G) towards position- and time-sensitive detectors at opposite ends of the 3-D momentum imaging spectrometer. The detectors consist of a Multi-Channel Plate (MCP) stack in chevron configuration with a delay-line anode readout~\cite{Roentdek,Jagutzki}. The electron and ion detectors are a three layer hex-anode with a 80~mm MCP stack and a two layer quad-anode with a 120~mm MCP stack, respectively, where a charge carrier's 3-D momentum is encoded into its hit position on the detector and its time-of-flight relative to the laser trigger. In this configuration, the energy resolution of the spectrometer for the electrons is roughly 1:10.

The laser system has been described previously \cite{Sturm}, so we provide only a brief overview that emphasizes some recent modifications. Near-infrared (NIR) laser pulses (12~mJ, 45~femtoseconds, 800~nm) from a 50~Hz Ti:sapphire laser system are frequency doubled using a 0.25~mm thick beta-barium borate (BBO) crystal, where the copropagating NIR and blue fields are then separated using two dichroic mirrors. The reflected 400~nm photons ($\sim$3.6~mJ, $\sim$50~fs) are used to generate femtosecond VUV pulses via HHG, by loosely focusing (f~=~6~m) the 400~nm pulses into a 10~cm long gas cell containing 2 Torr of krypton. The resulting VUV frequency comb is then separated from the 400~nm driving field by reflection from three Si mirrors near Brewster's angle for the blue light, resulting in a suppression of the driving field by a factor of $<10^{-6}$. The 3\textsuperscript{rd} harmonic (133~nm, 9.3~eV) is selected via transmission through a 0.25~mm thick MgF$_{2}$ window, which totally suppresses the 5\textsuperscript{th} harmonic and above and maintains the femtosecond pulse duration of the 3\textsuperscript{rd} harmonic. The residual 400~nm pulses are also separated from the 3\textsuperscript{rd} harmonic pulse temporally by roughly 700~fs, due to the difference in the Group Velocity Dispersion (GVD) of the window at $\omega_0$ and $3\omega_0$ \cite{Allison,Li}. After transmission through the window we estimate the pulse duration of the 3\textsuperscript{rd} harmonic to be $\sim$30~fs, based on its spectral bandwidth, its estimated attochirp, and the thickness and GVD of the MgF$_{2}$ window \cite{Sekikawa1,Sekikawa2}. The femtosecond 9.3~eV pulses are then back-focused (f~=~15~cm) into the 3-D momentum imaging spectrometer using a protected Al mirror, the reflectance of which has been measured to be 43\% at 9.3~eV \cite{Larsen}. The pulse energy of the 3\textsuperscript{rd} harmonic on target is approximately 10~nJ, which was measured using a pair of broadband VUV filters (Acton Optics FB130-B-1D.3) and a calibrated photodiode. From our previous NOTPSI measurement of argon \cite{Larsen2}, the VUV spectrum of the 9.3 eV femtosecond pulse is known to be roughly Gaussian in shape and possesses no sidebands or pedestal, characteristic of the VUV frequency combs generated in the single-color high harmonic generation.

A rotationally and vibrationally cold beam of nitrogen molecules (approximately 80~K) is prepared via an adiabatic expansion through a 0.03~mm nozzle, which is then collimated by a pair of skimmers. This molecular jet propagates perpendicular to the focused VUV beam, where the two intersect in the interaction region of the spectrometer ($\sim$0.01 $\times$ 0.01 $\times$ 0.20~mm), resulting in NOTPSI of the target molecules at an average rate of of $\sim$0.3 events per VUV pulse.  

\section{\label{sec:level3}Theory}

The potential energy curves of the neutral N$_2$ states relevant to this study were computed using the multi-reference configuration interaction (MRCI) capability of MOLPRO \cite{doi:10.1002/wcms.82, werner_molpro_2015}, with single- and double excitations from an active space including four $\sigma_g$, four $\sigma_u$, two $\pi_u$ and two $\pi_g$ orbitals, with the $1\sigma_g$ and $1\sigma_u$ core orbitals always doubly occupied. These orbitals were optimized using a state averaged complete active space self-consistent field (CASSCF) calculation with one state of each symmetry, $^1\Sigma_g^+$, $^1\Pi_u$, $^1\Sigma_u^-$, and $^1\Delta_u$, included. The one-electron basis set was aug-cc-pVTZ \cite{doi:10.1063/1.456153, doi:10.1063/1.462569}, augmented by four $s$, four $p$, and four $d$ Rydberg diffuse functions \cite{Kaufmann_1989} centered at the bond midpoint.   

\begin{figure}[h]
\begin{center}
\includegraphics[width=1.0\columnwidth,clip=false]{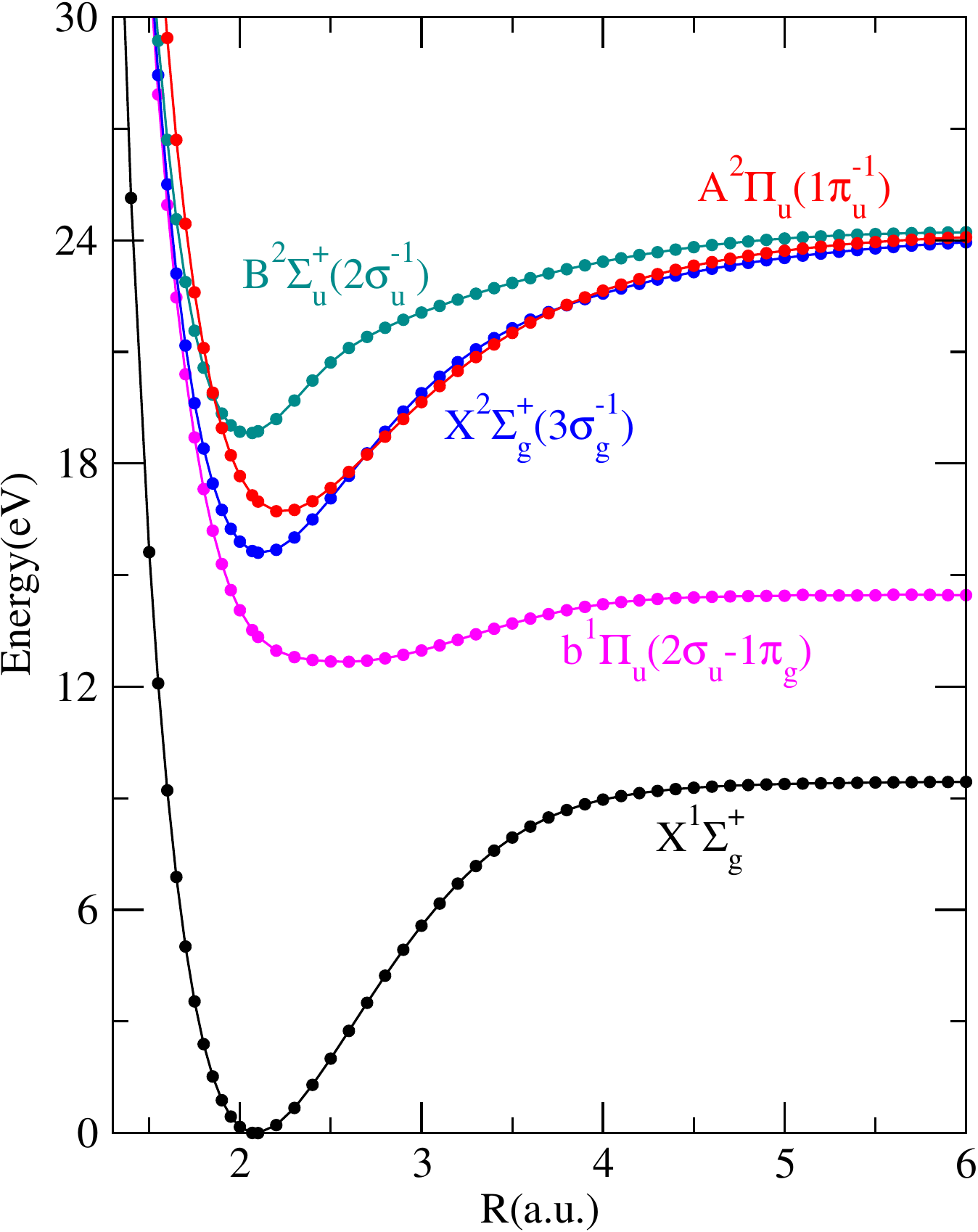} 
\caption{Relevant potential energy curves of N$_2$ and N$_2^+$. The figure shows the lowest $^1\Pi_u$ state of N$_2$ as well as the lowest three states of N$_2^+$.}
\label{fig:pecs}
\end{center}
\end{figure}

One-photon ionization cross sections and photoelectron $\beta$ parameters were calculated at a fixed internuclear distance $R=2.07$ a.u. using a multichannel configuration interaction method (MCCI) \cite{doi:10.1063/1.468841, doi:10.1063/1.471632}. The scattering state at a fixed energy is represented on a single-center expanded grid, with a partial wave expansion up to $l_{max}=80$. The full $N$-electron scattering state was then written as a close-coupling expansion containing a sum of the products of $(N-1)$-electron ion state wave functions multiplied with one-electron photoelectron wave functions. The lowest eight ion states were included in the close-coupling expansion. The initial and ion states were computed performing a numerical CAS-CI using the orbitals from the CASSCF discussed above. We considered as the intermediate state for ionization the b$^1\Pi_u$ ($2\sigma_u \rightarrow 1\pi_g$) valence state shown in Fig.~\ref{fig:pecs}, which is the first dipole allowed excited state of N$_2$ lying at $\sim$13.5 eV above the ground state \cite{CHAN199381, doi:10.1139/p69-072}. The three lowest ionization thresholds of N$_2$ were found to be $\sim$15.6 eV, $\sim$17.1 eV and $\sim$18.8 eV corresponding to ionization into the X$^2\Sigma_g^+$, A$^2\Pi_u$, and B$^2\Sigma_u^+$ N$_2^+$ ionic states, respectively. 

Two-photon single ionization $\beta_2$ and $\beta_4$ anisotropy parameters, have been calculated within the time-independent second order perturbation theory framework \cite{doi:10.1063/1.1896951, DIXON198421, doi:10.1063/1.442074, doi:10.1080/00268977600100931, doi:10.1063/1.444174}. For simplicity we have assumed a resonant transition to the b$^1\Pi_u$ excited state as the first photon absorption. Calculating the cross sections and anisotropy parameters for non-resonant two-photon single ionization of N$_2$ in lowest order perturbation theory requires a sum of single photon ionization amplitudes over a substantial number of intermediate states, which lies beyond the scope of this study. The dipole couplings between the ground state and the intermediate state have been calculated using MOLPRO.

\section{\label{sec:level4}Results}

\begin{figure}[h!]
\centering
    {
    \subfigure[]{%
        \includegraphics[width=8.0cm]{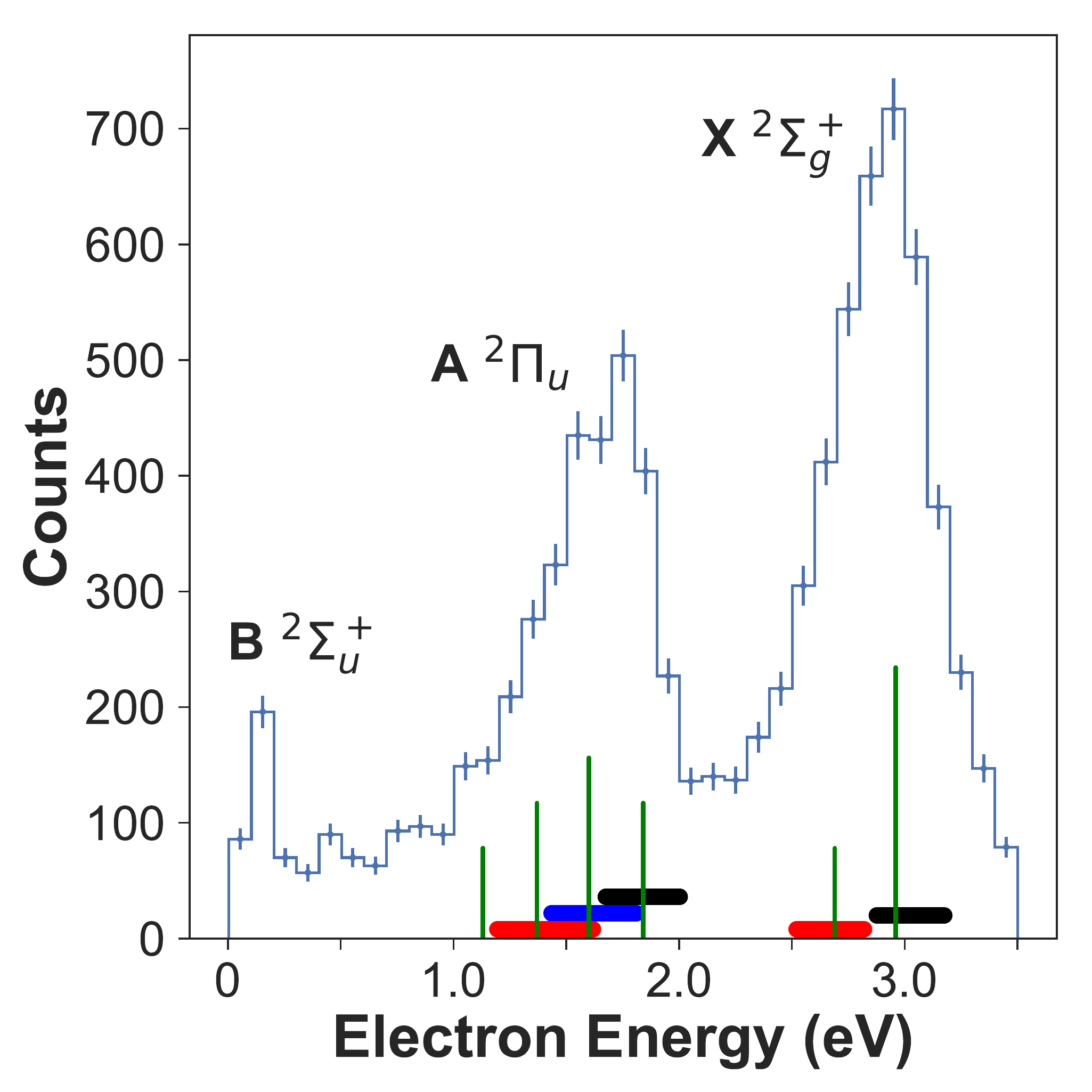}}}
    {
    \subfigure[]{%
        \includegraphics[width=8.0cm]{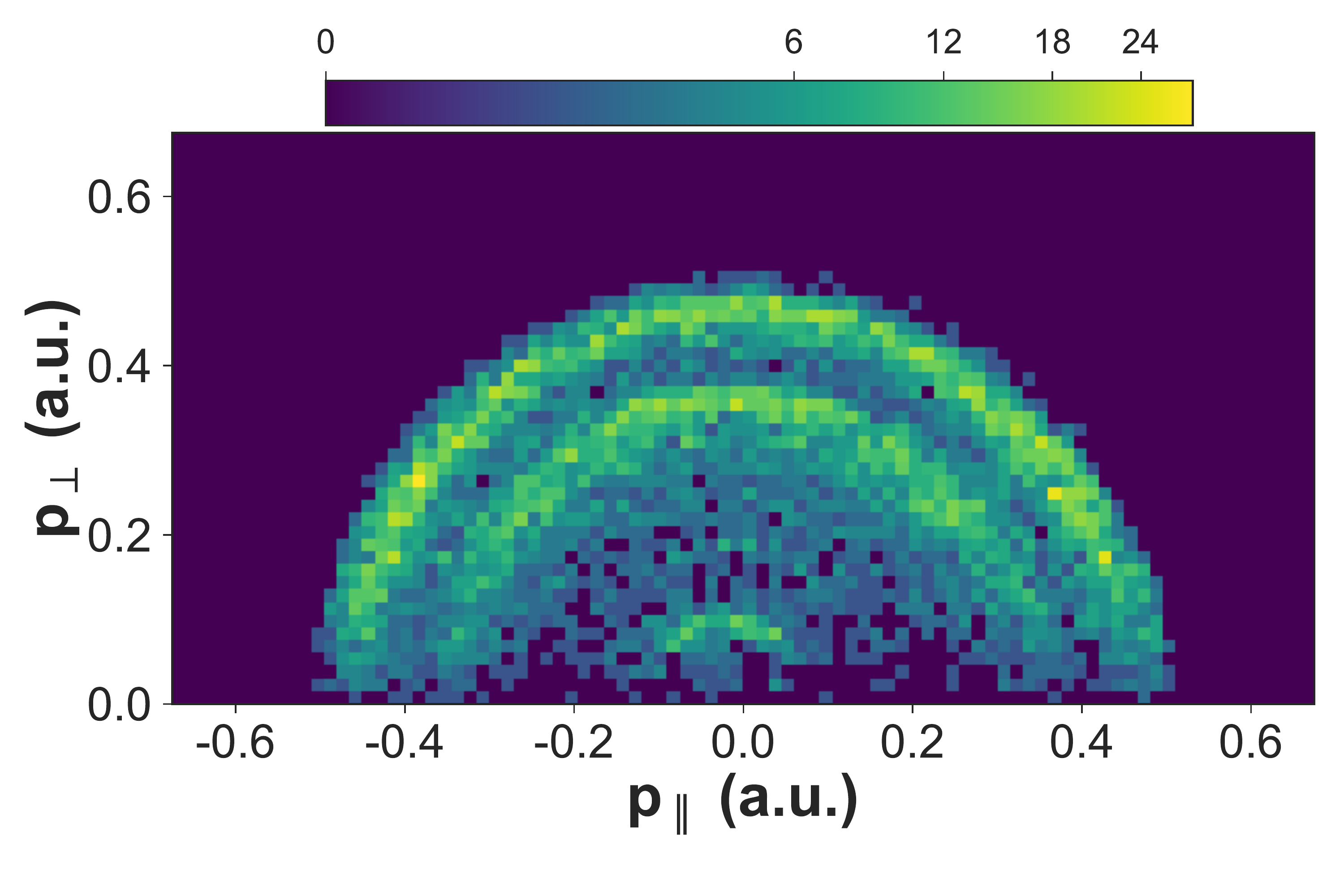}}}
\caption{(a) The measured photoelectron energy spectrum for NOTPSI of N$_2$. The colored horizontal lines indicate different energy regions of the X $^2\Sigma^+_g$ and A $^2\Pi_u$ states that are subsequently analyzed. The vertical green lines roughly indicate the location of the relevant vibrational levels of the two ionic states and their corresponding one-photon oscillator strengths. (b) Photoelectron momentum distribution parallel and perpendicular to the VUV polarization.}
\label{fig:Ee_p}
\end{figure}

The measured photoelectron energy spectrum is presented in Fig.~\ref{fig:Ee_p}(a), where we observe three peaks centered near 2.88~eV, 1.63~eV, and 0.15~eV, corresponding with the X $^2\Sigma^+_g$, A $^2\Pi_u$, and B $^2\Sigma^+_u$ ionic states of molecular nitrogen, respectively. The two-photon excitation energy and bandwidth is $\sim$18.6~eV and $\sim$300~meV, respectively. Hence, the measured linewidths of the three cationic states in Fig.~\ref{fig:Ee_p}(a) are largely determined by the spectral bandwidth of the VUV pulse, which exceeds the vibrational level spacing (see green vertical lines). We point out that although our central two-photon energy lies below the ionization threshold of the B $^2\Sigma^+_u$ ionic state, the two-photon bandwidth of the excitation pulse enables its lowest vibrational level to be accessed. The X $^2\Sigma^+_g$ and A $^2\Pi_u$ state peaks are partitioned into two and three different photoelectron kinetic energy regions, respectively, for subsequent analysis. The horizontal black and red lines in Fig.~\ref{fig:Ee_p}(a) indicate the high and low energy regions, respectively, while the horizontal blue line indicates a middle energy region that is considered in the case of the A $^2\Pi_u$ ionic state. The X $^2\Sigma^+_g$ state energy slices range from 2.50--2.85~eV and 2.85--3.20~eV, while the A $^2\Pi_u$ state slices span 1.18--1.64~eV, 1.41--1.83~ev, and 1.65--2.03~eV. These regions capture photoelectrons emerging from ionization events to different target state vibrational levels. The vertical green line segments in Fig.~\ref{fig:Ee_p}(a) indicate the energetic locations of the relevant vibrational levels of the X $^2\Sigma^+_g$ and A $^2\Pi_u$ ionic states and their corresponding FC factors~\cite{Yencha}. In the X $^2\Sigma^+_g$ state, the high energy region roughly corresponds with photoionization to the $\nu$=0 level of the target, and the low energy region captures photoionization to the $\nu$~=~1 level. In the A $^2\Pi_u$ state, the high energy slice roughly corresponds with populating the $\nu$~=~0, 1 vibrational levels of the target, the middle energy slice with the $\nu$~=~1, 2, 3 levels, and the low energy slice with the $\nu$~=~2, 3, 4 levels. Most of these slices sum over two or three vibrational levels, as the two-photon bandwidth, energy resolution, and statistics do not permit a fully vibrationally resolved measurement.

The photoelectron momentum distributions for the transverse and parallel momentum components with respect to the VUV polarization vector for the three ionic states are shown in Fig.~\ref{fig:Ee_p}(b). Here we see three arcs of differing photoelectron momentum, corresponding to each of the three populated molecular ionic states. To gain more insight into the photoelectron emission pattern for each of these three ionic states, we turn to the angle-differential photoionization cross section.

\begin{figure}[h!]
\centering
    {\subfigure[]{\includegraphics[width=5.25cm, trim=0.45cm 0.6cm 0.6cm 0.4cm, clip]{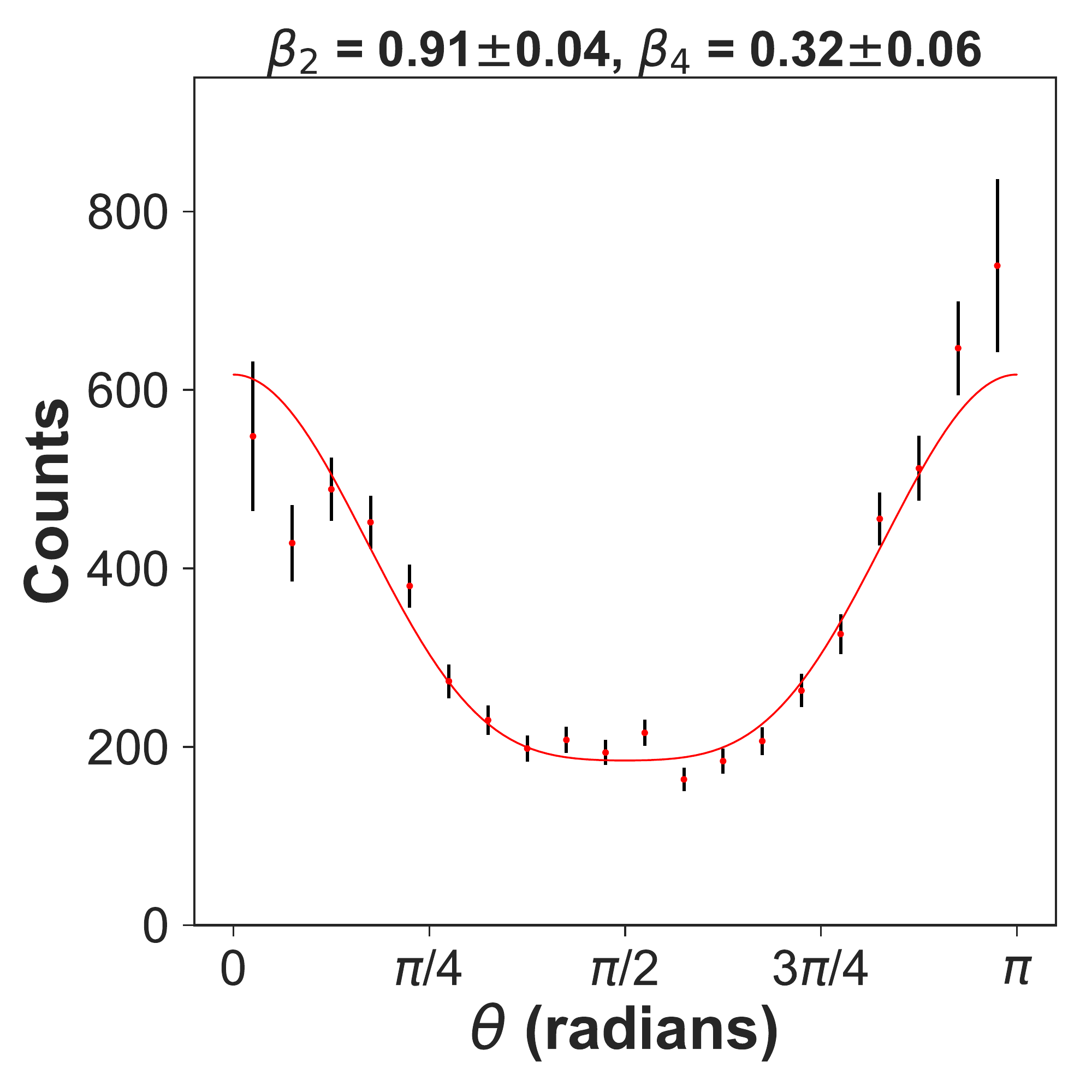}}}
    {\subfigure[]{\includegraphics[width=5.25cm, trim=0.45cm 0.6cm 0.6cm 0.4cm, clip]{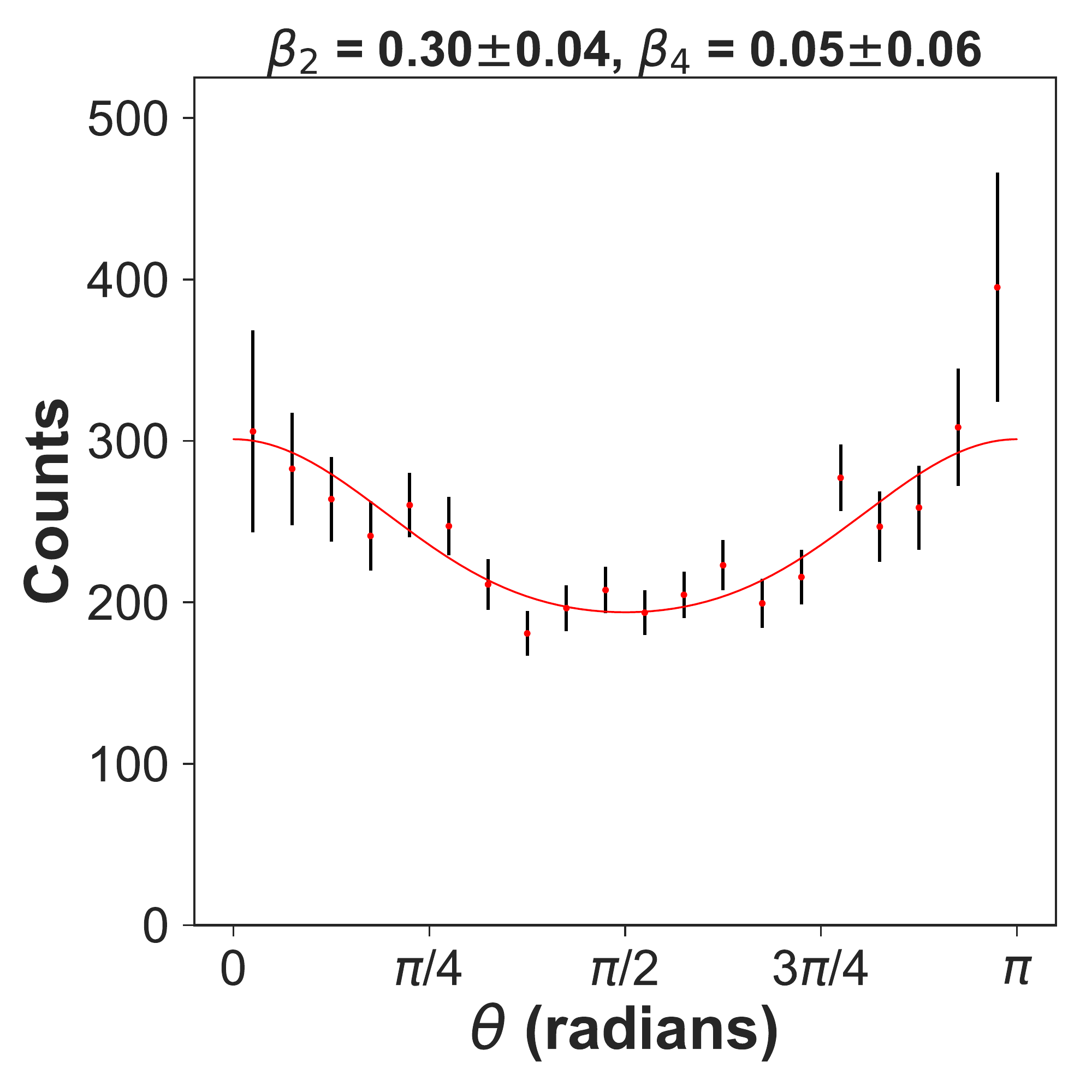}}}
    {\subfigure[]{\includegraphics[width=5.25cm, trim=0.45cm 0.6cm 0.6cm 0.4cm, clip]{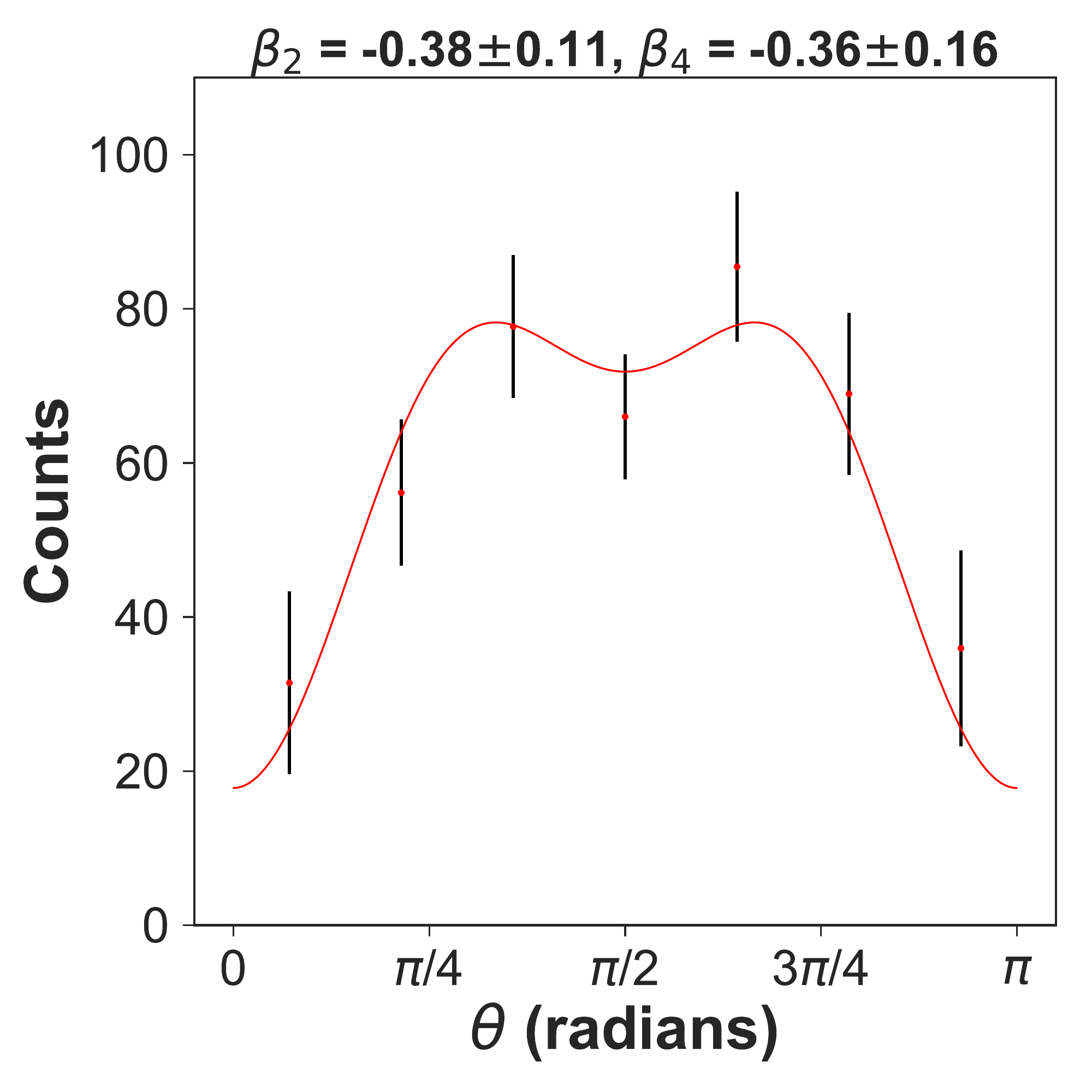}}}
\caption{The energy-integrated photoelectron angle-differential photoionization cross section for NOTPSI of N$_2$ to the first three ionic states, (a) the X $^2\Sigma^+_g$ state, (b) the A $^2\Pi_u$ state, and (c) the B $^2\Sigma^+_u$ state. The data are fitted using equation (1), where the retrieved $\beta$ parameters are displayed above each plot.}
\label{fig:theta}
\end{figure}

For two-photon ionization processes the energy-dependent angle-differential photoionization cross section of a target by linearly polarized light is given by
\begin{equation}
    \frac{d^2\sigma}{d\Omega dE} = \frac{\sigma_0(E)}{4\pi} [1 + \beta_{2}(E)P_{2}(\cos\theta) + \beta_{4}(E)P_{4}(\cos\theta)],
\label{eq:DDCS}
\end{equation}
where $\sigma_0$ is the total photoionization cross section, $\theta$ is the angle between the photoelectron momentum vector and the polarization vector of the light, E is the photoelectron kinetic energy, $\beta_{2}$ and $\beta_{4}$ are the second and fourth order anisotropy parameters, and $P_{2}$ and $P_{4}$ are the second and fourth order Legendre polynomials in variable $\cos\theta$. The measured energy-integrated angle-differential photoionization amplitude for each of the X $^2\Sigma^+_g$, A $^2\Pi_u$, and B $^2\Sigma^+_u$ ionic states are presented in Fig.~\ref{fig:theta}(a), (b), and (c), respectively. The data has been fitted (solid red line) in accordance with equation (1), using the projection method discussed in \cite{Liu}, while the error on the $\beta$ parameters was determined via statistical bootstrapping \cite{Efron}.

\begin{table}
\centering
\begin{tabular}{  c  c  c  } 
 \hline\hline
 State & $\beta_2$ & $\beta_4$ \\
 \hline 
 X $^2\Sigma^+_g$ & 0.91$\pm$0.04 & 0.32$\pm$0.06 \\
 A $^2\Pi_u$ & 0.30$\pm$0.04 & 0.05$\pm$0.06 \\
 B $^2\Sigma^+_u$ & -0.38$\pm$0.11 & -0.36$\pm$0.16 \\
 \hline
\end{tabular}
\caption{The anisotropy ($\beta$) parameters retrieved from fitting the energy-integrated photoelectron angular distributions shown in Fig.~\ref{fig:theta} according to equation (1) for the X $^2\Sigma^+_g$, A $^2\Pi_u$, and B $^2\Sigma^+_u$ states of N$_2^+$.}
\label{table:PADs_EI}
\end{table}

The PADs associated with the X $^2\Sigma^+_g$ and A $^2\Pi_u$ ionic states depicted in Fig.~\ref{fig:theta}(a) and (b) both exhibit peak intensity along the VUV polarization, more dramatically in the case of the X $^2\Sigma^+_g$ state, whereas the A $^2\Pi_u$ state PAD is more isotropic. The PAD associated with the B $^2\Sigma^+_u$ state in Fig.~\ref{fig:theta}(c) appears bimodal in shape with two photoelectron emission peaks roughly perpendicular to the VUV field, each slightly offset to opposite sides of $\pi/2$. The energy-integrated $\beta$ parameters associated with these three ionic states can all be found in Table~\ref{table:PADs_EI}.

\begin{figure}[h!]
\centering
    {\subfigure[]{\includegraphics[width=5.5cm, trim=0.45cm 0.6cm 0.6cm 0.4cm, clip]{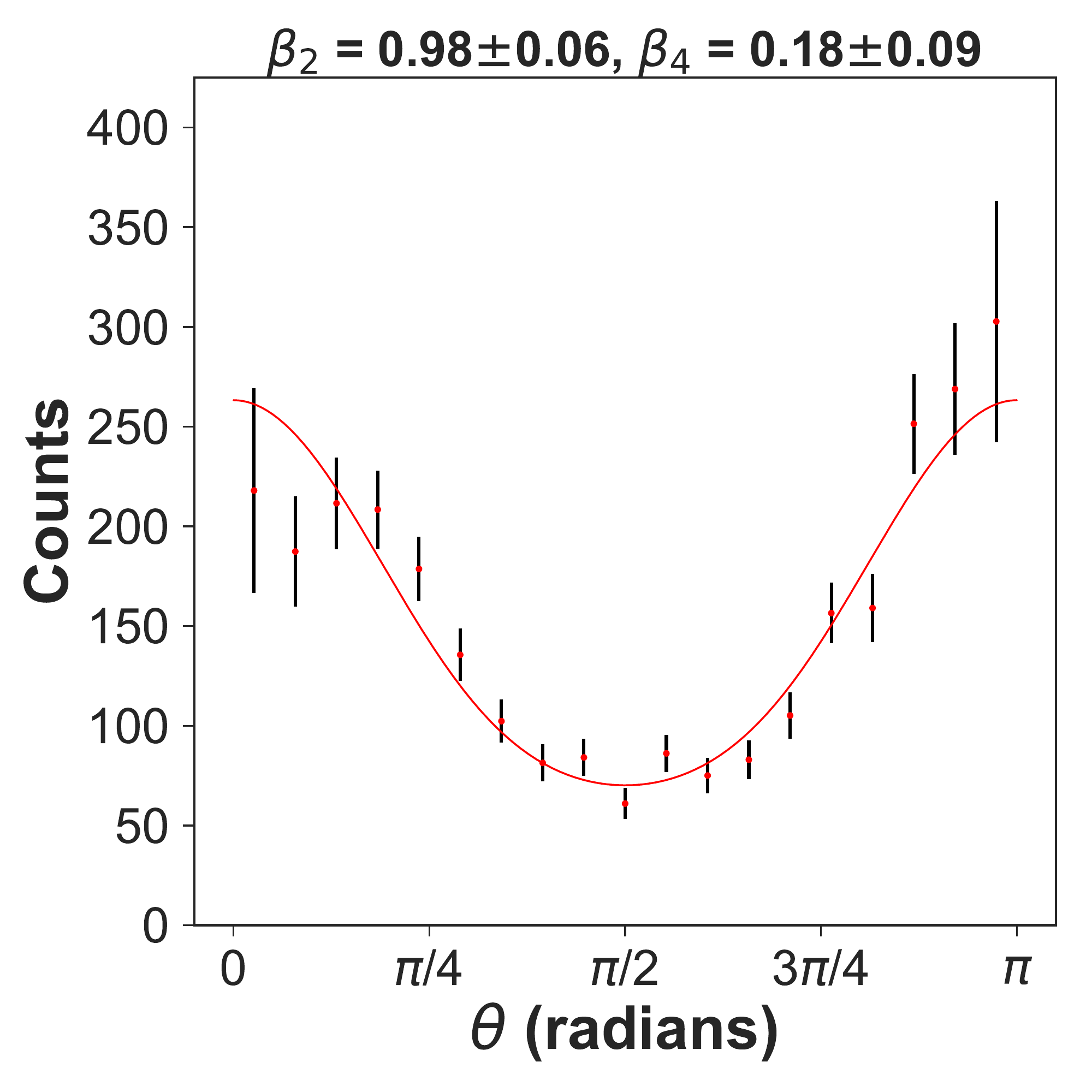}}}
    {\subfigure[]{\includegraphics[width=5.5cm, trim=0.45cm 0.6cm 0.6cm 0.4cm, clip]{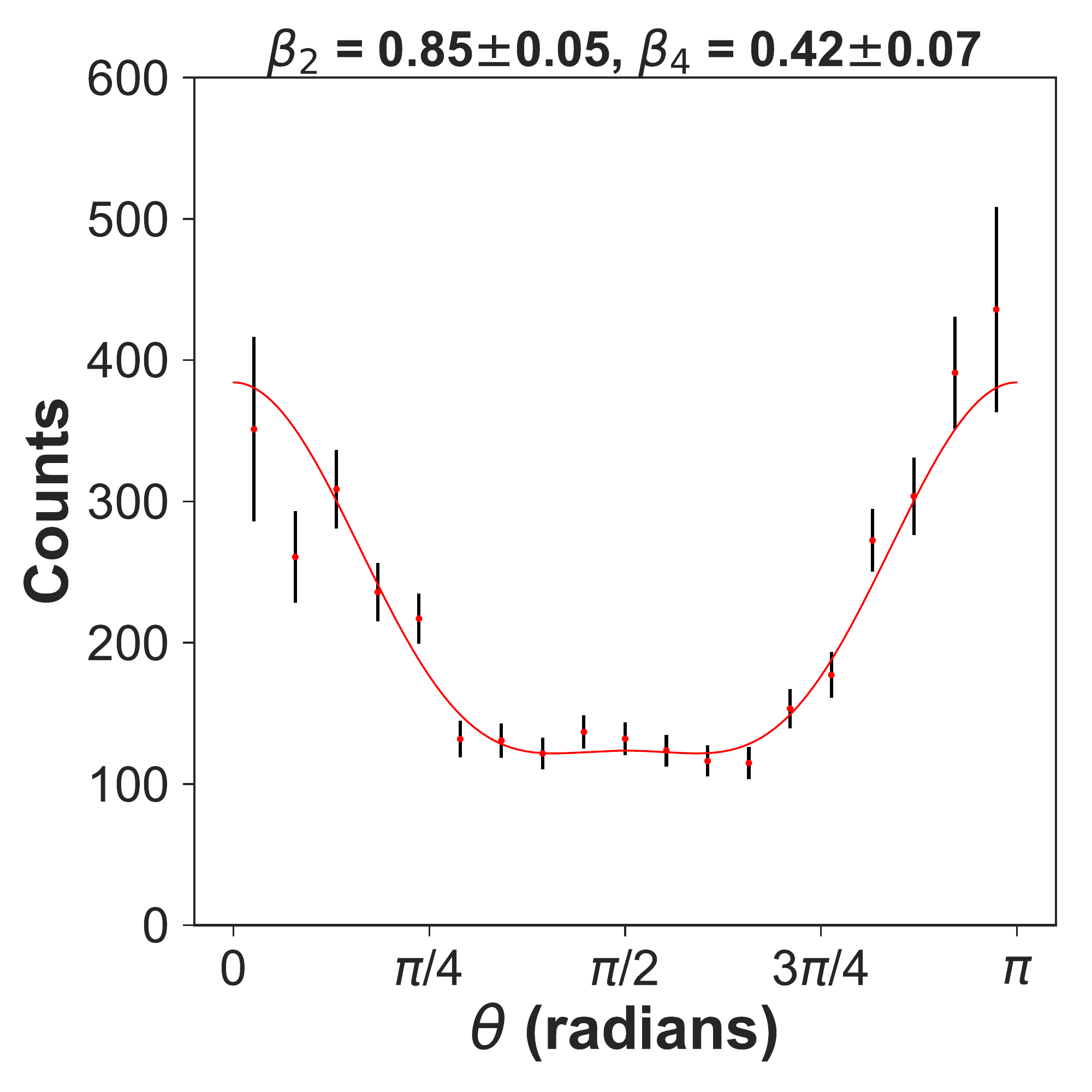}}}
\caption{The energy-resolved photoelectron angle-differential photoionization cross sections of N$_2$ following NOTPSI. Depicted is the photoelectron angular distribution for the X $^2\Sigma^+_g$ ionic state for two different energy regions in the photoelectron spectrum, indicated in Fig.~\ref{fig:Ee_p}, with the low kinetic energy region shown in (a) and the high energy region in (b). The data are fitted using equation (1), where the retrieved $\beta$ parameters are shown above each plot.}
\label{fig:ERtheta_X}
\end{figure}

In Fig.~\ref{fig:ERtheta_X} we show the energy-resolved PADs associated with the X $^2\Sigma^+_g$ ionic state, where we compare the low photoelectron energy portion with the high energy portion, indicated by the red and black horizontal lines in Fig.~\ref{fig:Ee_p}(a). The energy-resolved $\beta$ parameters for the X $^2\Sigma^+_g$ ionic state can be found in Table~\ref{table:PADs_ER}. We find that the PADs vary between the two regions, where $\beta_2$ undergoes a slight decrease in value, while $\beta_4$ increases by more than a factor of 2, as the photoelectron kinetic energy increases. Here, the high electron energy PAD in Fig.~\ref{fig:ERtheta_X}(b) exhibits a broader minimum perpendicular to the VUV polarization vector than the low energy PAD in Fig.~\ref{fig:ERtheta_X}(a).

\begin{figure}[h!]
\centering
    {\subfigure[]{\includegraphics[width=5.5cm, trim=0.45cm 0.6cm 0.6cm 0.4cm, clip]{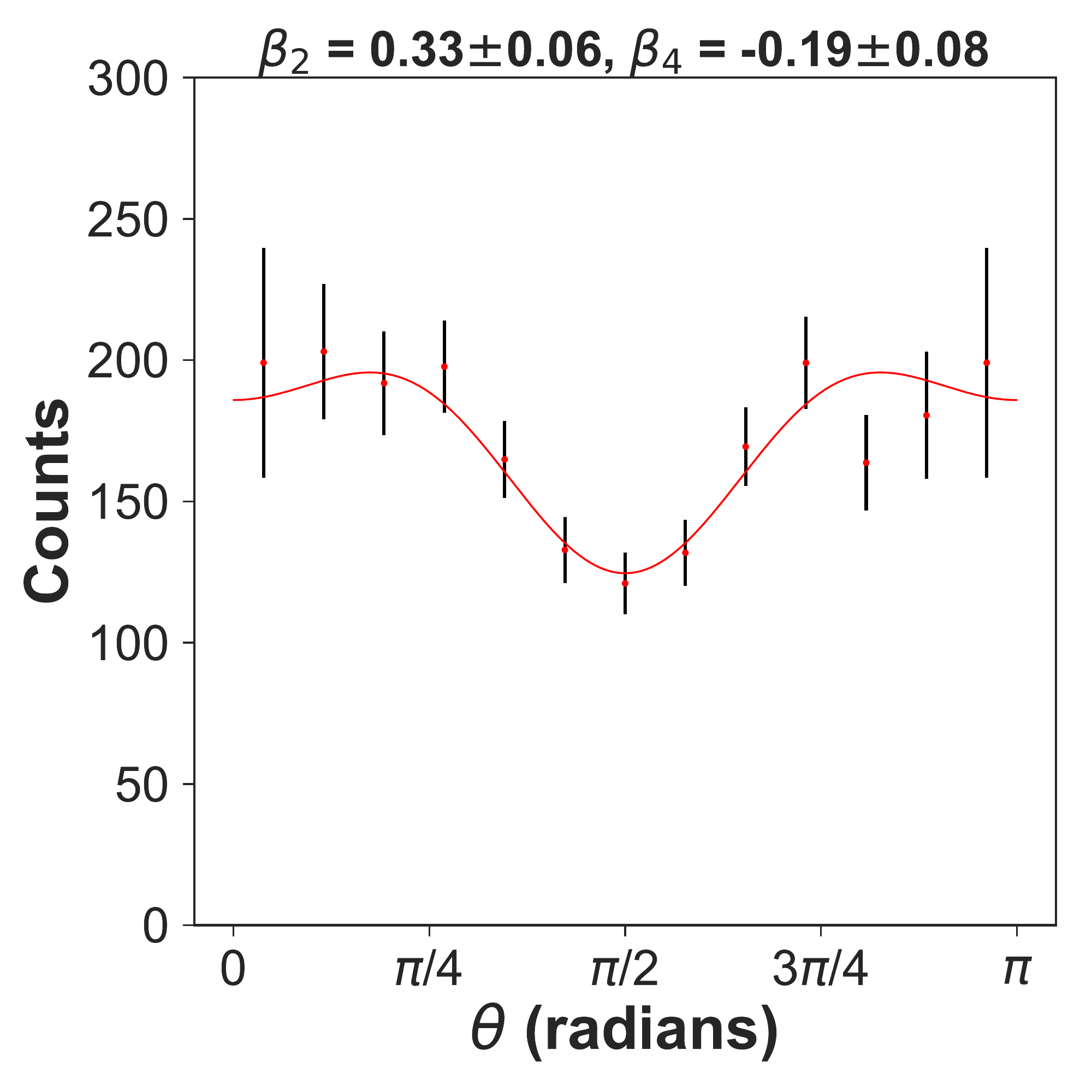}}}
    {\subfigure[]{\includegraphics[width=5.5cm, trim=0.45cm 0.6cm 0.6cm 0.4cm, clip]{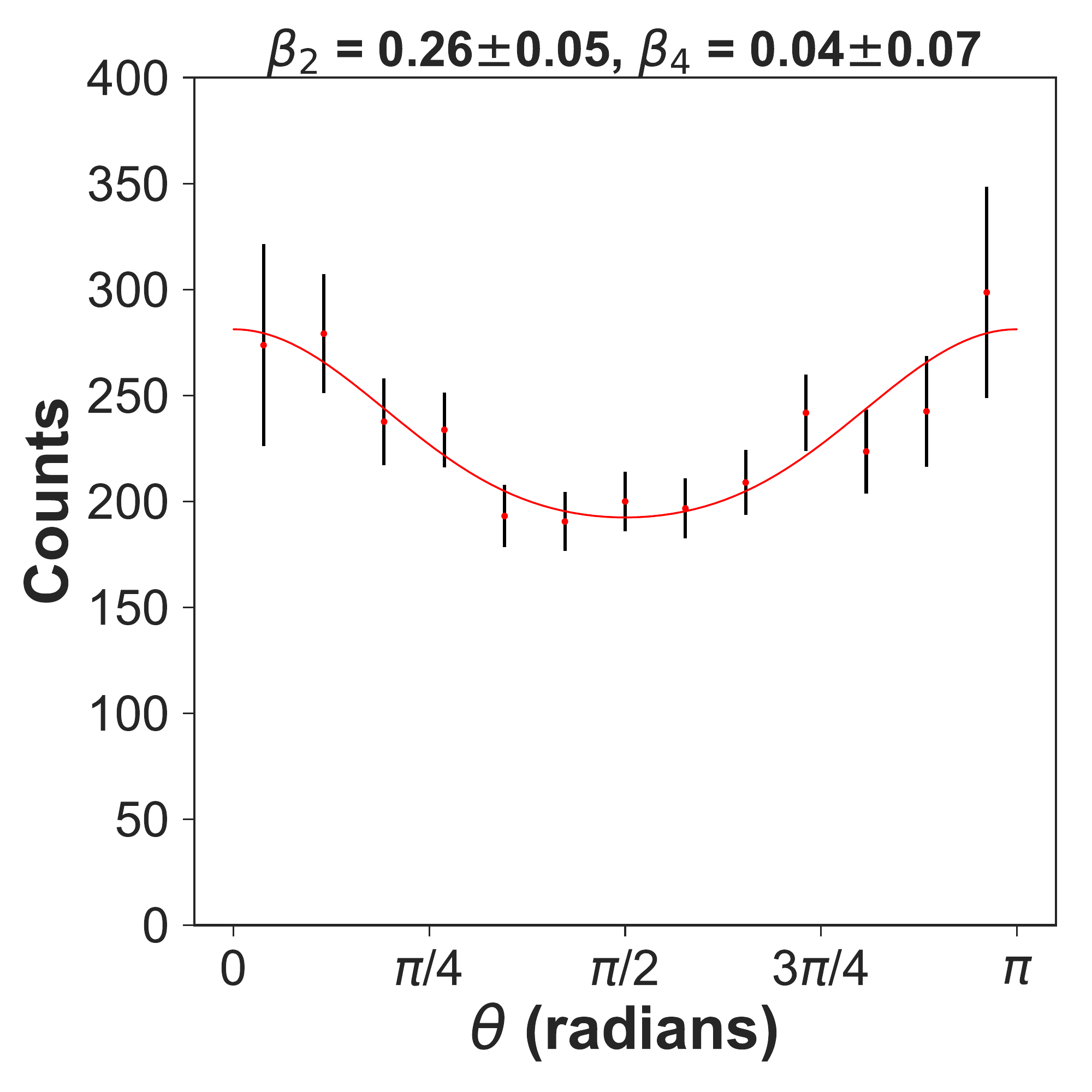}}}
    {\subfigure[]{\includegraphics[width=5.5cm, trim=0.45cm 0.6cm 0.6cm 0.4cm, clip]{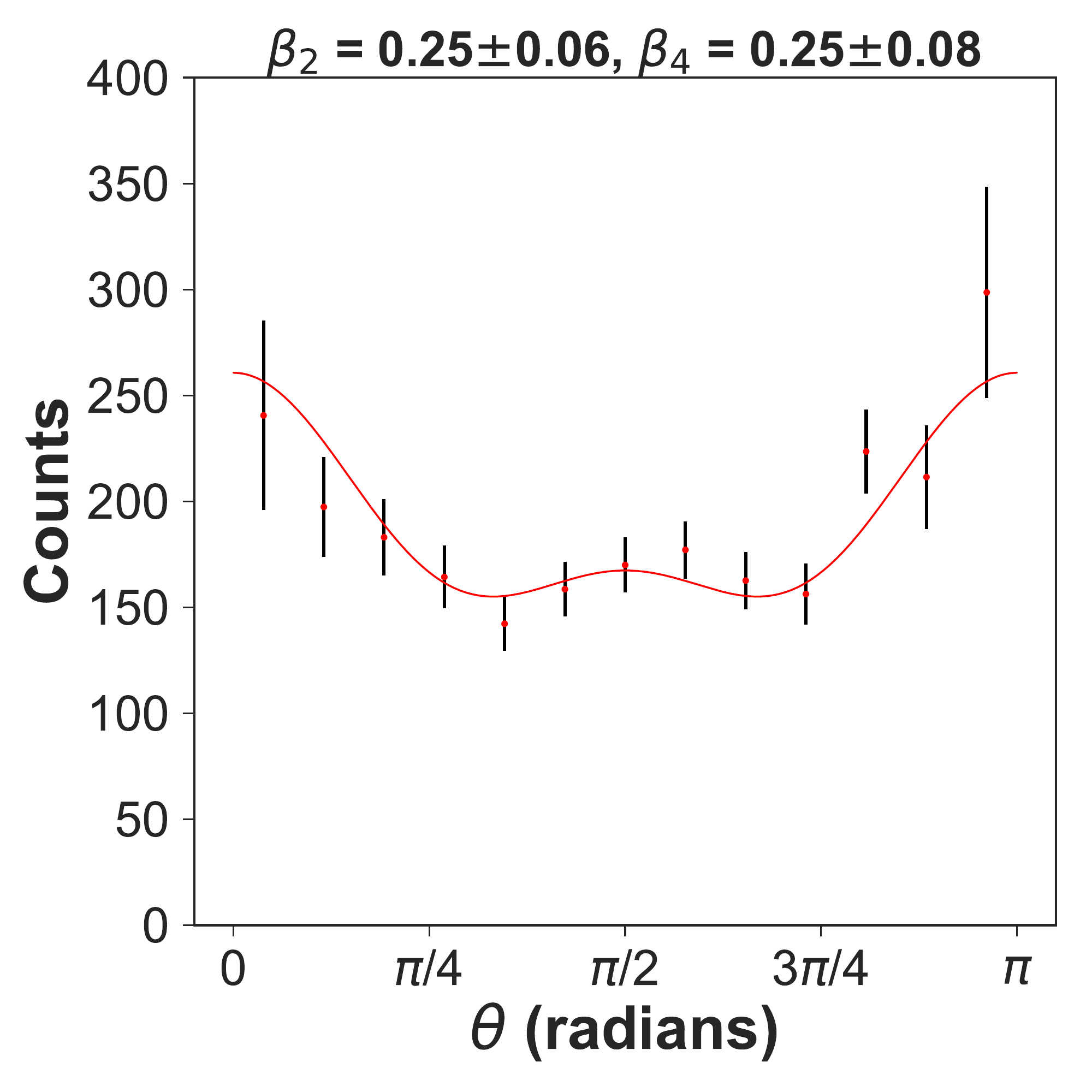}}}
\caption{The energy-resolved photoelectron angle-differential photoionization cross sections of N$_2$ following NOTPSI. Depicted is the photoelectron angular distribution for the A $^2\Pi_u$ ionic state for three different energy regions in the photoelectron spectrum, indicated in Fig.~\ref{fig:Ee_p}, with the low kinetic energy region shown in (a), the middle energy region in (b), and the high energy region in (c). The data are fitted using equation (1), where the retrieved $\beta$ parameters are shown above each plot.}
\label{fig:ERtheta_A}
\end{figure}

\begin{table}
\centering
\begin{tabular}{  c  c  c  c  } 
 \hline\hline
 State & Photoelectron Energy (eV) & $\beta_2$ & $\beta_4$ \\
 \hline
 \multirow{2}{2.9em}{X $^2\Sigma^+_g$} & 2.50--2.85 & 0.98$\pm$0.06 & 0.18$\pm$0.09 \\ 
  & 2.85--3.20 & 0.85$\pm$0.05 & 0.42$\pm$0.07 \\
 \hline
 \multirow{3}{2.9em}{A $^2\Pi_u$} & 1.18--1.64 & 0.33$\pm$0.06 & -0.19$\pm$0.08 \\
 & 1.41--1.83 & 0.26$\pm$0.05 & 0.04$\pm$0.07 \\
 & 1.65--2.03 & 0.25$\pm$0.06 & 0.25$\pm$0.08 \\
 \hline
\end{tabular}
\caption{The anisotropy ($\beta$) parameters retrieved from fitting the energy-resolved photoelectron angular distributions according to equation (1) for the X $^2\Sigma^+_g$ and A $^2\Pi_u$ ionic states of N$_2$, shown in Fig.~\ref{fig:ERtheta_X} and Fig.~\ref{fig:ERtheta_A}.}
\label{table:PADs_ER}
\end{table}

In Fig.~\ref{fig:ERtheta_A} we show the energy-resolved PADs associated with the A $^2\Pi_u$ ionic state, where we compare the low, middle and high energy regions, indicated by the red, blue and black horizontal lines in Fig.~\ref{fig:Ee_p}(a), respectively. The energy-resolved $\beta$ parameters associated with the A $^2\Pi_u$ ionic state can be found in Table~\ref{table:PADs_ER}. The PADs vary appreciably between the three slices, where $\beta_2$ slightly decreases in value, while $\beta_4$ increases in value dramatically and undergoes a sign change, as the photoelectron kinetic energy increases. The high energy PAD in Fig.~\ref{fig:ERtheta_A}(c) exhibits maxima parallel to the ionizing field and a smaller peak perpendicular to the field, while the low energy PAD in Fig.~\ref{fig:ERtheta_A}(a) shows a minimum perpendicular to the ionizing field and a smaller minimum parallel to the field. The middle energy PAD in Fig.~\ref{fig:ERtheta_A}(b) resembles the energy-integrated PAD of Fig.~\ref{fig:theta}(b), exhibiting a shallow minimum perpendicular to the VUV polarization vector.

For both states, the variation in the retrieved $\beta_2$ asymmetry parameter is not particularly significant, while the change in the $\beta_4$ parameter is more substantial. This is most striking in the A $^2\Pi_u$ state, where the value of $\beta_4$ changes sign, resulting in an appreciable variation in the PAD across the different photoelectron kinetic energy regions. Such a strong energy dependent variation in the PAD is unlikely to stem from the effect of the final vibrational level on the scattering wave function, as the two-photon bandwidth exceeds the spacing of the vibrational levels in the cationic states, where the two-photon bandwidth is $\sim$300~meV, while the level spacing in the A $^2\Pi_u$ ionic state is $\sim$230~meV \cite{Yencha}. Further, the different photoelectron energy regions considered here average over a family of adjacent vibrational levels, which should wash out the effect of the vibrational level on the photoelectron scattering wave function (i.e. our energy-resolved PADs integrate over numerous levels). Previous theoretical one-photon studies on the effect of the vibrational level of the target state of the photoelectron asymmetry parameter have indicated that the effect is negligible in the A $^2\Pi_u$ state \cite{Plesiat}, where we observe the strongest variation in the PAD. These effects are also unlikely to emerge from the shape resonance in the photoionization continuum of N$_2$ alone, as its peak is far higher in photoelectron kinetic energies ($\sim$32~eV) than the kinetic energies of the electrons corresponding to the populated continuum states in the present work.

\begin{figure}[h!]
\begin{center}
\includegraphics[width=1.0\columnwidth,clip=false]{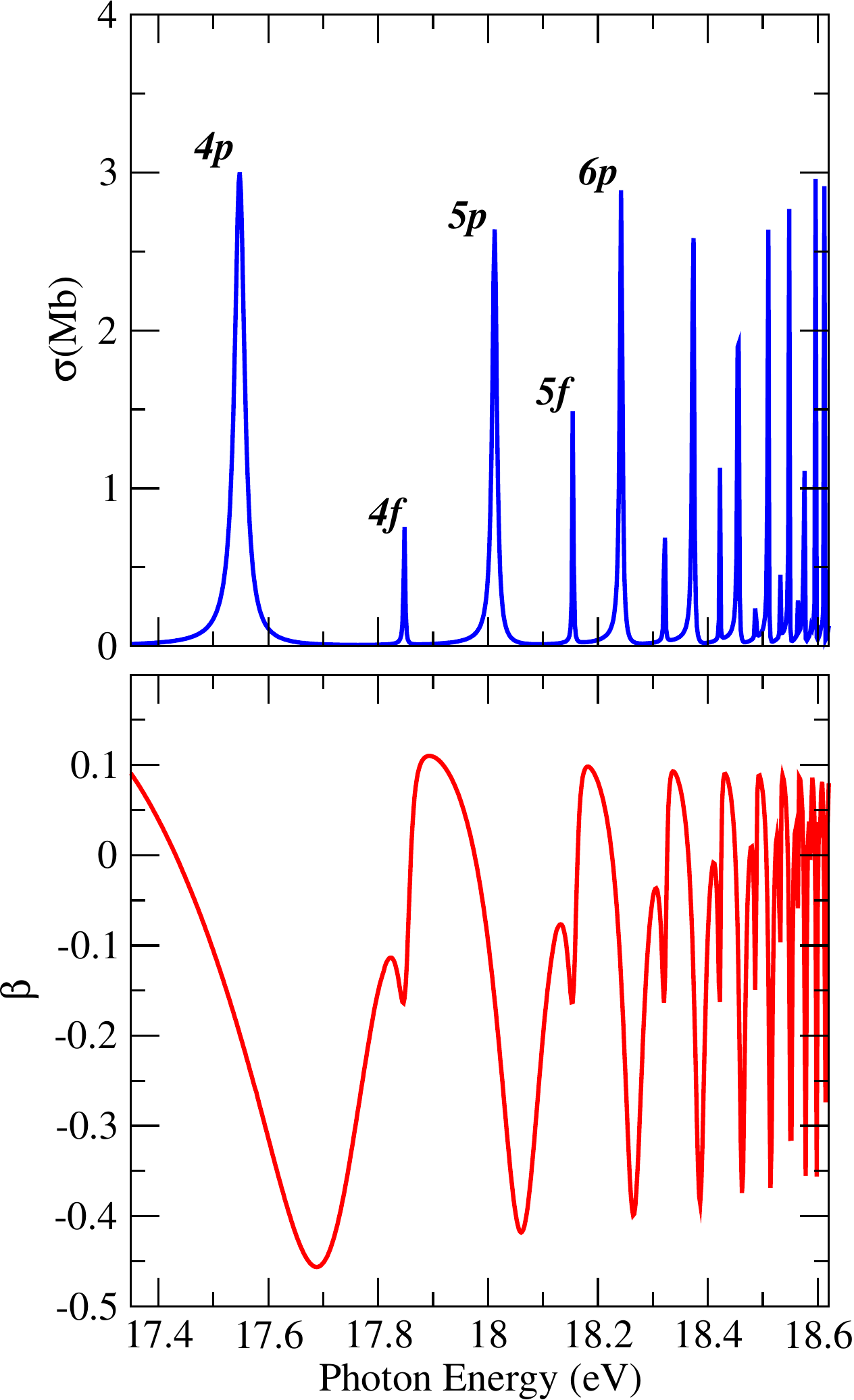}
\caption{Upper panel: Total photoionization cross section from the N$_2$ (b$^1\Pi_u$) state between the second (A$^2\Pi_u$) and third (B$^2\Sigma_u^+$) ionization thresholds, while leaving the system in $^1\Sigma_g^+$ final symmetry. The $nl$ labels indicate the $np\sigma_u$ and $nf$($\delta_u$ or $\sigma_u$) series of the autoionizing states converging to the N$_2^+$ (B$^2\Sigma_u^+$) cation state. Lower panel: Corresponding photoelectron asymmetry parameter for the one photon transition. Note that the abscissa in this figure and Figs.~\ref{fig:cross_section_Pg}, \ref{fig:cross_section_Dg}, and \ref{fig:beta_24} represents the energy of a two-photon transition from the ground state, where one photon is on resonance with the intermediate b$^1\Pi_u$ state, while the second photon energy is varied.}
\label{fig:cross_section_Sg}
\end{center}
\end{figure}

\begin{figure}[h!]
\begin{center}
\includegraphics[width=1.0\columnwidth,clip=false]{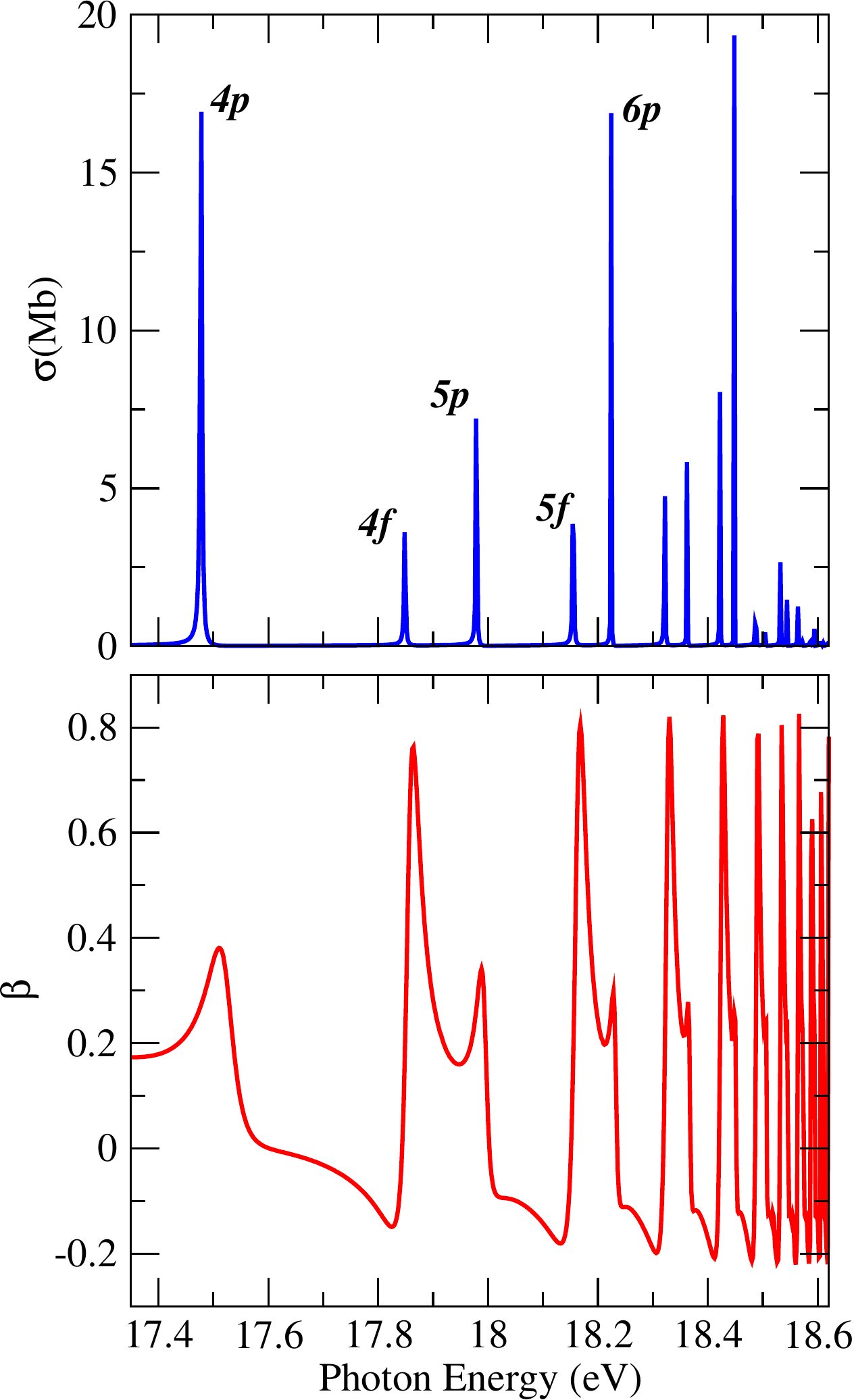}
\caption{As in Fig.~\ref{fig:cross_section_Sg} but leaving the system in  $^1\Pi_g$ final symmetry. The $nl$ labels indicate the $np\pi_u$ and $nf$($\phi_u$ or $\pi_u$) series of the autoionizing states converging to the N$_2^+$ (B$^2\Sigma_u^+$) cation state.}
\label{fig:cross_section_Pg}
\end{center}
\end{figure}

\begin{figure}[h!]
\begin{center}
\includegraphics[width=1.0\columnwidth,clip=false]{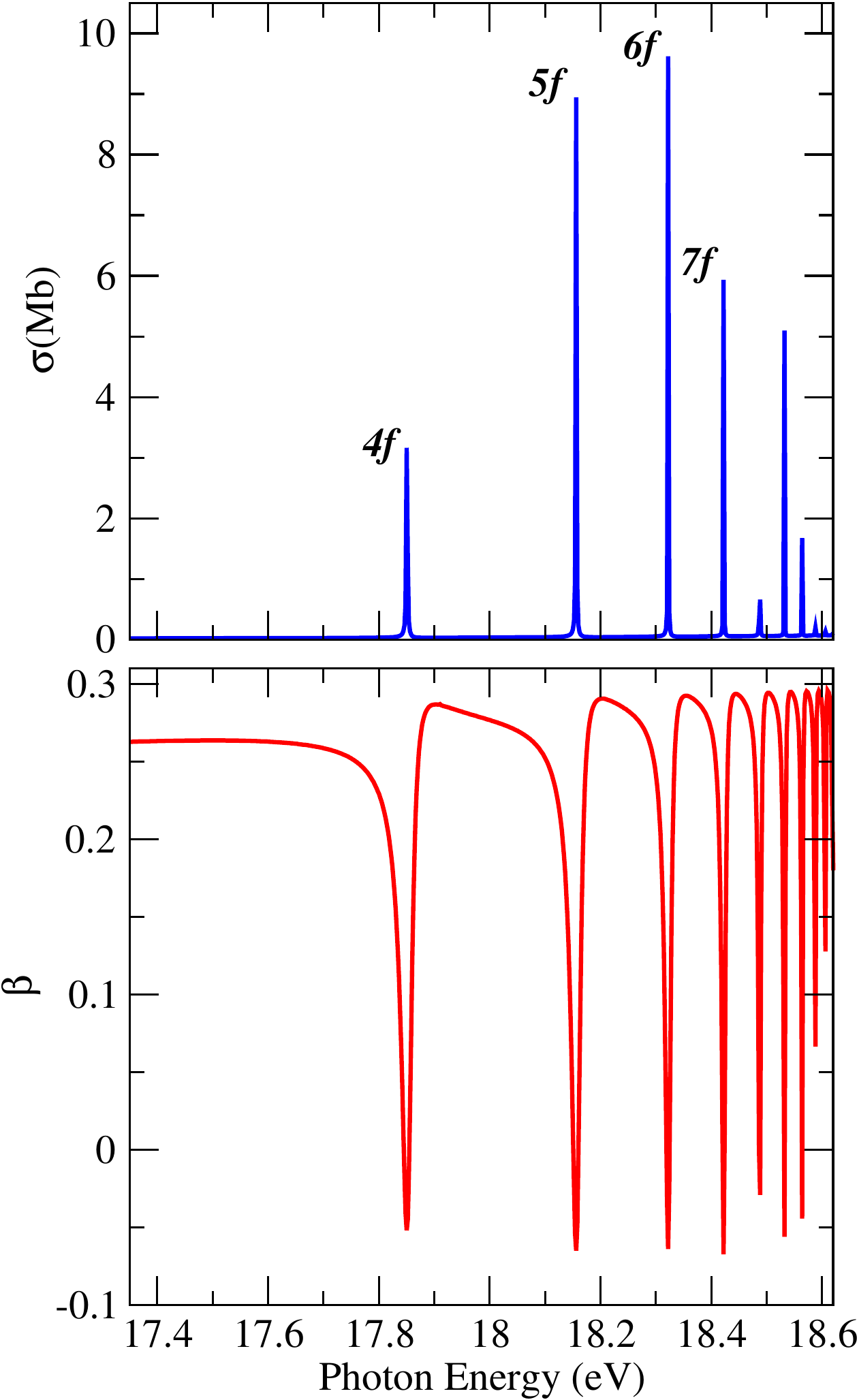}
\caption{As in Fig.~\ref{fig:cross_section_Sg} but leaving the system in $^1\Delta_g$ final symmetry. The $nl$ labels indicate the $nf$($\delta_u$ or $\sigma_u$) series of the autoionizing states converging to the N$_2^+$ (B$^2\Sigma_u^+$) cation state.}
\label{fig:cross_section_Dg}
\end{center}
\end{figure}

In order to better understand the experimental findings, in particular the rapid evolution in the measured PADs that are associated with the A$^2\Pi_u$ ionic state, we have performed one-photon single ionization calculations from a selected excited state of neutral N$_2$. We chose the b$^1\Pi_u$ state, which is the lowest dipole-allowed electronic state lying $\sim$13.5 eV above the N$_2$ ground-state. The total one-photon ionization cross sections from the N$_2$ (b$^1\Pi_u$) state along with the corresponding photoelectron $\beta$ parameters for the $^1\Sigma_g^+$,  $^1\Pi_g$, and $^1\Delta_g$ dipole-allowed total symmetries are shown in Figs.~\ref{fig:cross_section_Sg}, \ref{fig:cross_section_Pg}, and \ref{fig:cross_section_Dg}, respectively. 
All these symmetries have to be taken into account, as our experiment probes randomly oriented molecules, and the two-photon energy lies below the dissociative ionization threshold, preventing the molecular orientation at the instant of photoionization from being determined without prealignment. The cross sections were calculated at a single fixed nuclear geometry over an energy range between 17.4 eV and 18.7 eV, where the X$^2\Sigma_g^+$ and A$^2\Pi_u$ ionic states are accessible (see Fig.~\ref{fig:pecs}).

The photoionization cross section in each symmetry exhibits a collection of structures that can be attributed to the Hopfield series of autoionizing states converging to the B$^2\Sigma_u^+$ state of the ion. The Hopfield series has been previously studied, theoretically and experimentally, in the one-photon ionization from the N$_2$ ground state leading to $^1\Sigma_u^+$ and $^1\Pi_u$ final symmetries. In contrast, the corresponding photoionization cross sections from an excited state of N$_2$ have not been reported theoretically to the best of our knowledge. Two different series, $np$ and $nf$, have been assigned in the cross sections using the Rydberg equation  
\begin{equation}
E_n=(IP)-\frac{R}{(n-\delta)^2},
\label{eq:MQDT}
\end{equation}
where $E_n$ represents the energy of the autoionizing state, $R$ is the Rydberg constant, $n$ is the principal quantum number of the Rydberg electron, and $\delta$ is the quantum defect, which depends on the angular momentum $l$. For $l=0$ a large quantum defect, i.e. $\delta \sim 1$, is expected while $\delta$ decreases rapidly as $l$ increases such that $nd$ and $nf$ series ($l=2,3$) are characterized by $\delta \sim 0$.
 
While the $np$ and $nf$ series coexist in the $^1\Sigma_g^+$ and $^1\Pi_g$ symmetries, seen in the upper panel of Fig.~\ref{fig:cross_section_Sg} and \ref{fig:cross_section_Pg}, only the $nf$ is visible in the $^1\Delta_g$ symmetry, seen in the upper panel of Fig.~\ref{fig:cross_section_Dg}. The $np$ series, $\sigma_u$ or $\pi_u$, are characterized by a slightly different quantum defect, $\delta \sim 0.56$ or $\delta \sim 0.66$, respectively, appearing thus at different energy positions. In contrast, the $nf$ series presents the same quantum defect number $\delta \sim -0.002$ in each final symmetry. The effect of the autoionizing states on the PADs is highly significant. The $\beta$ parameter dramatically changes as the photon energy crosses their energy position, seen in the lower panels of Fig.~\ref{fig:cross_section_Sg},  \ref{fig:cross_section_Pg}, and \ref{fig:cross_section_Dg}. In addition, both series of autoionizing states generate completely different values of the $\beta$ parameter. For instance as observed in Fig.~\ref{fig:cross_section_Pg} the $5f$ autoionizing state gives rise to an asymmetry parameter of $\beta=0.81$, while the $6p$ state is characterized by $\beta=0.30$. Is important to mention that although the magnitudes of the cross sections and $\beta$ parameters depend on the initial state, the positions of these same series of autoionizing states do not change when a different N$_2$ excited state is chosen as the initial state for the calculation.      

\begin{figure}[h!]
\begin{center}
\includegraphics[width=1.0\columnwidth,clip=false]{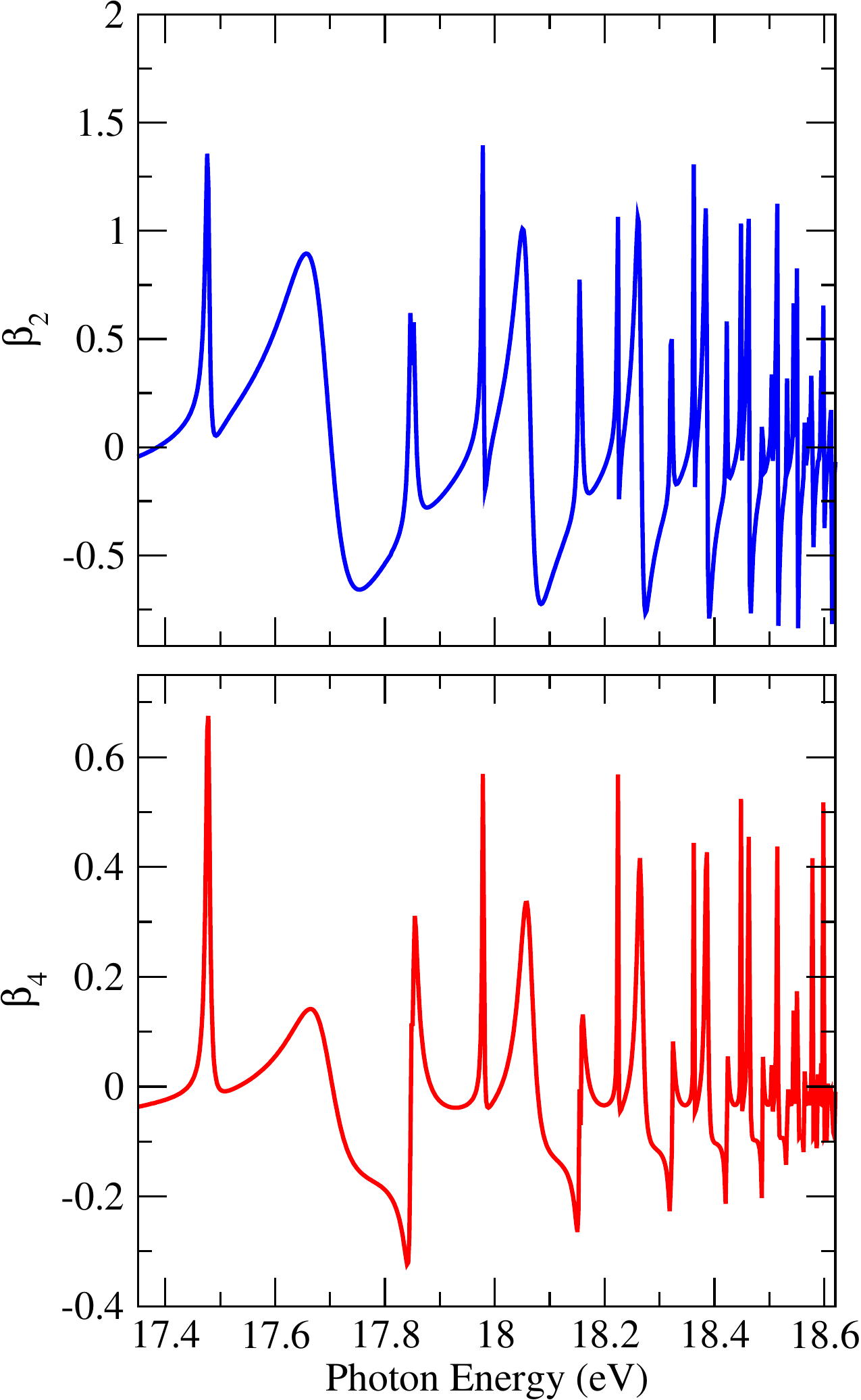} 
\caption{Two-photon photoelectron asymmetry parameters, $\beta_2$ (upper panel) and $\beta_4$ (lower panel), correlated to the A$^2\Pi_u$ state of the N$_2^+$ ion.}
\label{fig:beta_24}
\end{center}
\end{figure}

An accurate calculation of the cross section for non-resonant two-photon single ionization of N$_2$ in lowest order perturbation theory would require a sum of single photon ionization  amplitudes over a large number of ungerade intermediate electronic states, which is well beyond the scope of the present study. To simplify the calculation, while providing further insight into the experimental observations, we have performed two-photon ionization calculations using the neutral b$^1\Pi_u$ state as the single intermediate state. Fig.~\ref{fig:beta_24} shows the calculated $\beta_2$ and $\beta_4$ parameters characterizing the two-photon ionization into the A$^2\Pi_u$ state of the ion. Both parameters exhibit strong variations as a function of the photon energy, as observed previously in Figs.~\ref{fig:cross_section_Sg}, \ref{fig:cross_section_Pg}, and \ref{fig:cross_section_Dg}. This further underlines the effect of the autoionizing states on the PADs, as the autoionizing states appearing in each of the possible final symmetries are now taken into account. Considering the selected photoelectron kinetic energy intervals (see Fig.~\ref{fig:Ee_p}(a)), the $np$ and $nf$ Rydberg series contribute differently depending on the respective kinetic energy region, which results in completely different PADs as depicted in panels (a) to (c) of Fig.~\ref{fig:ERtheta_A}. The finite bandwidth of the VUV pulses used in the present experiments produce photoelectrons from several final ion vibrational levels. It has also been shown in previous vibrationally-resolved single-channel calculations, which do not treat autoionization processes~\cite{Plesiat}, that the influence of vibrational levels of the final state on the asymmetry parameter for the A$^2\Pi_u$ state are negligible. This prediction is consistent with the present calculations, stressing the major role of the continuum autoionizing states in NOTPSI of N$_2$, which manifests in a rapid variation in the anisoptropy parameters. Going beyond the scope of the present work, a quantitative comparison between the theoretical and experimental PADs would however require further calculations that take into account the non-resonant character of the first photon absorption step. This implies including dipole couplings between the ground state and several excited intermediate states as well as incorporating continuum-continuum dipole couplings.  

\section{\label{sec:level5}Conclusion}

We have presented energy- and angle-resolved experimental and theoretical results on the photoionization dynamics of N$_2$, analyzing ionic-state-selective photoelectron angular distribution following NOTPSI of neutral nitrogen molecules using coincidence 3-D momentum imaging and intense 9.3~eV femtosecond pulses produced via 400~nm driven HHG. Two-photon single ionization populates the first three ionic states of the molecular nitrogen cation, where we find that the PADs associated with the X $^2\Sigma^+_g$ and A $^2\Pi_u$ ionic states of N$_2^+$ both vary with minor changes in the photoelectron kinetic energy. We attribute this strong variation in the electron emission patterns on the photoelectron energy to the excitation and decay of dipole-forbidden autoionizing resonances of different total symmetry that belong to the Hopfield series. In addition to the direct ionization channel, these continuum-embedded discrete states provide an indirect pathway to the X $^2\Sigma^+_g$ and A $^2\Pi_u$ final states. The 9.3~eV photon energy used in the present experiment can be very slightly detuned using plasma blueshifting of the harmonics, however this effect is small and limited in range. This unfortunately prevented a careful and thorough exploration of the effect of the photon energy on the photoionization dynamics in the current experimental set-up. However, a more tunable VUV light source, such as a monochromatized VUV free-electron laser, could allow the photon energy to be scanned and the effect of intermediate resonances to be explored in more detail in the future.

Despite our limited theoretical treatment, only invoking a single intermediate state and hence treating the two-photon ionization as resonant, our calculations provide insight into the non-resonant two-photon ionization dynamics, highlighting the important role of the dipole-forbidden autoionizing states on shaping the PADs and resulting $\beta$ parameters. Further calculations involving a larger number of intermediate states lie beyond the scope of this work, however, the computed energies of the dipole-forbidden autoionizing states are insensitive to the intermediate state chosen, which suggests that such autoionizing resonances will remain important in the non-resonant regime.

The two competing pathways of autoionization and direct ionization can interfere for electrons ionized to the same final state. Experiments sensitive to such interference phenomena require the energy and angular momentum of detected autoionization electrons and photoelectrons (corresponding to indirect and direct pathways, respectively) to be identical. As the electron in the continuum can share angular momentum with the molecular axis in several ways, interference can only be observed for fixed-in-space molecular N$_2^+$ cations. As mentioned before, in the present work the target N$_2$ molecules were randomly orientated, and hence we could not isolate the small subset of events that would sensitively exhibit interference effects in the PADs. Future experiments could employ a multi-color interrogation scheme that enables prealigning the molecular axis via an additional laser pulse, which is able to impulsively align the neutral molecule with respect to the polarization vector of the intense VUV beam prior to ionization. The necessary degree of alignment to achieve molecular orientations where the final total wave function for the two ionization pathways will be indistinguishable is hereby expected to depend on the density of states of the selected target molecule.

\section{\label{sec:level6}Acknowledgments}

Work at LBNL was supported by the U.S. Department of Energy Office of Basic Energy Sciences, Division of Chemical Sciences, Biosciences, and Geosciences under contract No. DE-AC02-05CH11231 and used resources of the National Energy Research Computing Center, a United States Department of Energy (US DOE) Office of Science User Facility supported by the Office of Science of the DOE, and the Lawrencium computational cluster resource provided by the IT Division at the Lawrence Berkeley National Laboratory.

\bibliography{N2Refs}

\end{document}